\documentclass[a4paper,fleqn,usenatbib]{mnras}

\pdfoutput=1

\usepackage{txfonts}

\usepackage[T1]{fontenc}
\usepackage{ae,aecompl}

\usepackage{graphicx}	
\usepackage{amssymb}	
\usepackage[flushleft]{threeparttable} 

\title[Simulating a slow bar in UGC 628]{Simulating a slow bar in the low surface brightness galaxy UGC 628}

\author[M. H. Chequers et al.]{
Matthew H. Chequers,$^{1}$\thanks{E-mail: mchequers@astro.queensu.ca}
Kristine Spekkens,$^{1,2}$
Lawrence M. Widrow$^{1}$
\newauthor \, and Colleen Gilhuly$^{1}$
\\
$^{1}$Department of Physics, Engineering Physics, and Astronomy, Queen's University, Kingston, ON K7L 3N6, Canada\\
$^{2}$Department of Physics, Royal Military College of Canada, P.O. Box 17000, Station Forces, Kingston, ON K7L 7B4, Canada
}

\date{Accepted XXX. Received YYY; in original form ZZZ}

\pubyear{2016}

\begin{document}
\label{firstpage}
\pagerange{\pageref{firstpage}--\pageref{lastpage}}
\maketitle

\begin{abstract}
We present a disc-halo N-body model of the low surface brightness
galaxy UGC 628, one of the few systems that harbours a ``slow" bar with
a ratio of corotation radius to bar length of $\mathcal{R} \equiv
R_c/a_b \sim 2$. We select our initial conditions using SDSS DR10
photometry, a physically motivated radially variable mass-to-light
ratio profile, and rotation curve data from the literature. A global
bar instability grows in our submaximal disc model, and the disc
morphology and dynamics agree broadly with the photometry and
kinematics of UGC~628 at times between peak bar strength and the onset
of buckling. Prior to bar formation, the disc and halo contribute
roughly equally to the potential in the galaxy's inner region, giving
the disc enough self gravity for bar modes to grow. After bar
formation there is significant mass redistribution, creating a
baryon dominated inner and dark matter dominated outer disc. This
implies that, unlike most other low surface brightness galaxies, UGC~628 is not dark matter dominated everywhere. Our model nonetheless implies that
UGC~628 falls on same the relationship between dark matter fraction
and rotation velocity found for high surface brightness galaxies, and
lends credence to the argument that the disc mass fraction measured at
the location where its contribution to the potential peaks is not a
reliable indicator of its dynamical importance at all radii.
\end{abstract}




\section{Introduction}
\label{sec:intro}

Bars provide a testing ground for our theories of galaxy evolution,
structure and dynamics.  In particular, since the susceptibility of a
galactic disc to bar-like instabilities is a function of the
relative contributions to the gravitational potential of the disc,
bulge, and dark halo, bars can be used to constrain mass models and
break the disc-halo degeneracy that plagues rotation curve
decomposition. The strength, pattern speed, and length of a bar all
depend on the structure of the host galaxy, a point illustrated in
numerous N-body experiments.  For example, the quintessential strong,
thin, and long bar can develop through an $m$=2 instability in galaxy
models where the disc contribution to the centripetal force is
comparable to that of the bulge and halo inside a few disc scale
lengths \citep[e.g see the review by][]{sellwood2014}.  By contrast, if the disc completely dominates the
centripetal force in the inner region, the bar that develops will be
shorter, fatter, and more boxy \citep{athanassoula2003}.
\\
\indent
The contribution of the disc to the centripetal force can be
represented by the dimensionless ratio

\begin{equation} \label{eq:discmaximality}
\mathcal{F}_{X} = V_d\left (XR_d\right)/V\left (XR_d\right ),
\end{equation}

\noindent where $V\left (R\right )$ is the total circular speed at
radius $R$, $V_d$ is the contribution to $V$ from the
disc, and $X$ denotes a multiple of the exponential disc scale length $R_d$ \citep{vanalbadaetal1985,sackett1997}. $\mathcal{F}_{X}$ is commonly evaluated at $R = 2.2R_d$, where the disc contribution to the mass distribution peaks. It is often more useful to consider disc maximality in terms of $\mathcal{F}_{X}^{2}$, rather than the classical quantity defined in equation~(\ref{eq:discmaximality}), since $\mathcal{F}_{X}^{2}$ is a more direct measure of the mass contribution from the disc to the total mass budget of the galaxy enclosed within the radius $X R_d$.  By convention, discs are described as maximal if $\mathcal{F}_{2.2}^{2} > 0.72$.  The first case described above corresponds to a submaximal disc ($\mathcal{F}_{2.2}^{2}\simeq 0.5$) while the second, a maximal one.
\\
\indent
In this paper, we consider the low surface brightness (LSB) galaxy UGC
628.  This galaxy has a clear photometric bar
\citep{debloketal2001,deblokbosma2002,cheminhernandez2009} and is
commonly classified as Sbc--Sm \citep{devaucouleursetal1991}. LSB galaxies are thought to be dark
matter dominated in the inner disc
\citep{bothunetal1997,deblokmcgaugh1997,debloketal2001,combes2002,deblokbosma2002,kuziodenarayetal2008} and indeed, the mass model for
UGC 628 by \citet{deblokbosma2002} has $\mathcal{F}_{2.2}^{2}\simeq$ 0.09--0.16.
The mere presence of a bar in a galaxy with such a small $\mathcal{F}_{2.2}^{2}$ value already
challenges our understanding of bar formation because self-gravity in the
disc is so low \citep{mayerwadsley2004}. Indeed, the bar fraction in LSB galaxies is a mere $\sim 4\%$ \citep{mihosetal1997}, while that in their high surface brightness counterparts is about an order of magnitude higher \cite[e.g.][]{marinovajogee2007}.
\\
\indent
Our main interest in UGC 628 is due to the claim by
\citet{cheminhernandez2009} that its bar is ``slow''.  A bar with
pattern speed $\Omega_p$ has a corotation resonance at radius $R_c$
defined by the condition $\Omega\left (R_c\right ) = V(R_c)/R_c =
\Omega_p$.  Corotation sets a theoretical upper bound on the length of
the bar $a_B$ and therefore the dimensionless ratio $\mathcal{R}
\equiv R_c/a_B$ is expected to be greater than unity.  Moreover, for a
given $a_B$, $\Omega_p<\Omega(a_B)$ so long as $\Omega$ is a
decreasing function of $R$, which is almost always the case.  Thus,
bars with $\mathcal{R}$ close to unity have a pattern speed as fast as
nature will allow.  By convention, bars are defined respectively as
``fast'' or ``slow'' depending on whether $\mathcal{R}$ is less than
or greater than 1.4.
\\
\indent
It is notoriously difficult to measure the length and pattern speed of
bars in real galaxies (and for that matter, simulated ones) and
therefore estimates of $\mathcal{R}$ are generally plagued by large
uncertainties.  Nevertheless, of the tens of galaxies \citep{rautiainenetal2008,aguerrietal2015} for which $\mathcal{R}$ has now been measured, the vast majority appear to be fast.  Indeed, UGC 628 is one of only three galaxies that are observed to have a slow bar, the others being the blue compact dwarf NGC 2915 \citep{bureauetal1999} and the dwarf irregular NGC 3741 \citep{banerjeeetal2013}.
\\
\indent
The preponderance of fast bars especially among massive, bright
galaxies is easy to understand.  Once a bar-like perturbation develops,
it causes circular orbits inside corotation to become elongated in the
same sense as the perturbation thereby enhancing the putative bar.
By contrast, circular orbits outside corotation are elongated
perpendicular to the perturbation \citep[See][]{contopoulos1980}.  The
bar rapidly develops and grows out to its corotation radius.  The bar
pattern speed may decrease with time, via dynamical friction \citep{chandrasekhar1943,mulder1983,weinberg1985} for example, however,
this only allows for more stars to participate in the bar mode as
corotation is pushed to larger radii.  Thus, $\Omega_p$ can decrease but the bar remains ``fast" \citep{athanassoula2013}.
\\
\indent
These arguments may not hold for LSB galaxies.  \citet{marinovajogee2007} showed that $a_B$ rarely exceeds $R_{25}$, the radius the
surface brightness isophote equals $25~{\rm mag~arcsec}^{-2}$.
Evidently, once $R_c$ approaches $R_{25}$ discs lack the surface
density out to corotation necessary to support a fast bar. Additionally, the strong shear in this region implied by flat galaxy rotation curves destabilizes the development of precessing bar orbits. Thus, it may not be surprising that an LSB galaxy (where the surface brightness is everywhere lower than in typical discs) like UGC 628 harbours one of the few known examples of a slow bar.
\\
\indent
The previous discussion suggests that $\mathcal{R}$ should depend on a
galaxy's morphological type.  This hypothesis is supported by
\cite{rautiainenetal2008} who estimated pattern speeds for 38 barred galaxies by modelling near-infrared and optical images. In short, they simulated the response of a gas and stellar disc
to a rigidly rotating $m$=2 potential perturbation and varied $\Omega_p$ until the simulated disc morphologies matched observations. They found that $\mathcal{R}$ gradually increased from early to late type galaxies; Sa through Sb galaxies
tended to have fast bars while Sbc through Scd galaxies tended to have
slow ones. By contrast, the bar pattern speeds measured by \citet{aguerrietal2015}, who applied the Tremaine-Weinberg
method \citep{tremaineweinberg1984} to stellar absorption line
maps and optical images of 15 CALIFA survey
galaxies, all point to fast bars with no dependence on morphology.
However, they did not include galaxies with Hubble type later than Sbc. Thus, whether bars are generally fast or slow remains an open question.
\\
\indent
In this paper, we present a dynamical N-body model for UGC 628 by
evolving an (initially) exponential stellar disc in a live
Navarro-Frenk-White (NFW) halo \citep{navarroetal1996} for $\sim$12 Gyr. Our goals are two-fold. The
first is to construct a model of this LSB galaxy that forms a bar.
Indeed, the exponential disc inferred by \citet{deblokbosma2002} for
UGC 628 has so little mass that, when evolved as an N-body model, it
develops flocculent spiral structure but no bar. We re-examine the
surface photometry of UGC 628 using updated multi-colour images and a
variable stellar mass-to-light ratio. We find a steeper surface
density profile than previously reported, from which we construct a
bar-unstable N-body model that reproduces the morphology and
kinematics of UGC 628 at late times. With this model in hand, we then
proceed to examine how the bar in UGC 628 redistributes mass in the
disc and the implications of that re-distribution for the dark matter
mass fraction in this LSB galaxy.
\\
\indent
Our procedure for constructing
the dynamical model of UGC 628 is outlined in
Section~\ref{sec:modelling}. This model provides initial conditions for our N-body simulations, which are described in Section~\ref{sec:results} where we compare our model to observations of UGC 628 and discuss
the quantitative properties of the bar. We discuss the implications of our results for the mass distribution in UGC 628 and summarize in Section~\ref{sec:discussion}. Throughout, we adopt a distance for UGC 628 of 71.2 Mpc and an inclination of 56\textsuperscript{o} \citep{cheminhernandez2009}.

\section{Modelling UGC 628}
\label{sec:modelling}

In this section we present our approach for constructing N-body
models of UGC 628. We derive the galaxy's surface density profile from
extant SDSS data and colour-mass-to-light ratio relations in Section~\ref{sec:modellingobservations} and
use it together with rotation curve data described in Section~\ref{sec:modellingrotationcurve} to generate
initial conditions for the simulations detailed in Section~\ref{sec:modellingnbody}.

\begin{figure} \centering
\includegraphics[angle=0,scale=0.26]{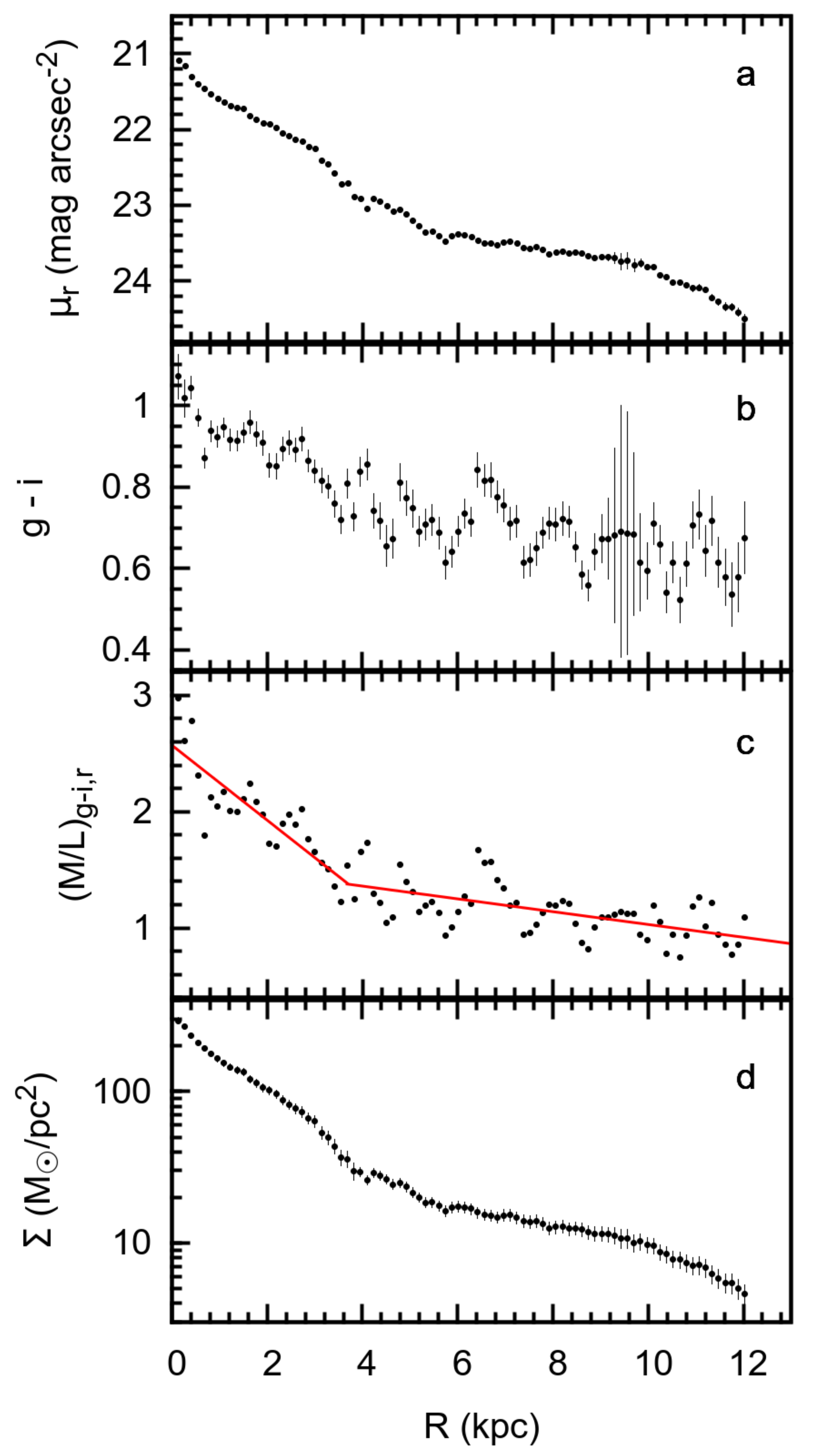}
\caption{Derivation of the stellar surface density profile of UGC 628. Panel a: SDSS DR10 $r$-band surface brightness profile. Panel b: $g$-$i$ colour profile derived from the $r$-band isophotal solution. Panel c: Mass-to-light profile derived from the \citet{intoportinari2013} ($g$-$i$,$r$) colour-band relation. The piecewise linear fit is shown as a solid red line. Panel d: Surface density profile produced by multiplying the surface brightness profile in panel a and the piecewise linear fit in panel c.  \label{fig:observationdata}}
\end{figure}

\subsection{Surface Density Profile from Photometry}
\label{sec:modellingobservations}
Fig.~\ref{fig:observationdata} describes how we infer the surface density profile of UGC 628 from photometric data. The image processing suite XVISTA\footnote{http://ganymede.nmsu.edu/holtz/xvista} was used to fit isophotal contours to an SDSS DR10 \citep{sdssdr10} $r$-band image. The break in the resulting $r$-band isophotal profile (Fig.~\ref{fig:observationdata}a) at $\sim$4 kpc divides the disc into two regions, with the inner region corresponding to the bar. The $r$-band solution was imposed on $g$- and $i$-band images and surface brightness profiles were generated in each band.
\\
\indent
\citet{intoportinari2013} tabulated colour-mass-to-light relations for a variety bands, and mass-to-light profiles derived using different combinations of $g$, $i$, and $r$ for UGC 628 are comparable. The $g$-$i$ colour profile presented in Fig.~\ref{fig:observationdata}b was selected to take advantage of the colour stability due to the largest difference between bands. The $g$-$i$ colour profile shows a number of features with amplitudes $\sim$~0.2~mag and length scales of order 1 kpc.  The most pronounced of these occur for $R$ between 3 and 9 kpc and are likely due to colour variations in the bar and spiral arms.  We note that the bar is also responsible for large, rapid variations in the ellipticity and position angle of the $r$-band isophotal contours, which were used to derive the $g$- and $i$-band surface brightness profiles. The inner portion of the resulting mass-to-light ratio profile, shown in Fig.~\ref{fig:observationdata}c, has higher mass-to-light and a steeper gradient than the outer portion of the profile. Thus, we model the mass-to-light ratio profile as a continuous piecewise linear function with two segments.
\\
\indent
Fig.~\ref{fig:observationdata}d shows the inferred surface density profile that was constructed by combining the $r$-band surface brightness profile in Fig.~\ref{fig:observationdata}a and the mass-to-light profile fit in Fig.~\ref{fig:observationdata}c. The uncertainties were calculated by propagating uncertainty from the $r$-band surface brightness profile and the piecewise linear fit of the mass-to-light profile. This surface density profile guided the selection of initial disc parameters in the N-body simulations of UGC 628, and serves as a basis of comparison with simulation snapshots. In comparison, the surface density profile of the \citet{deblokbosma2002} model of UGC 628 possesses a shallower slope and much smaller central density, which results from measuring an exponential scale length in the outer part of the surface brightness profile \citep{debloketal1995} and applying a constant mass-to-light ratio.

\begin{figure} \centering
\includegraphics[angle=0,scale=0.26]{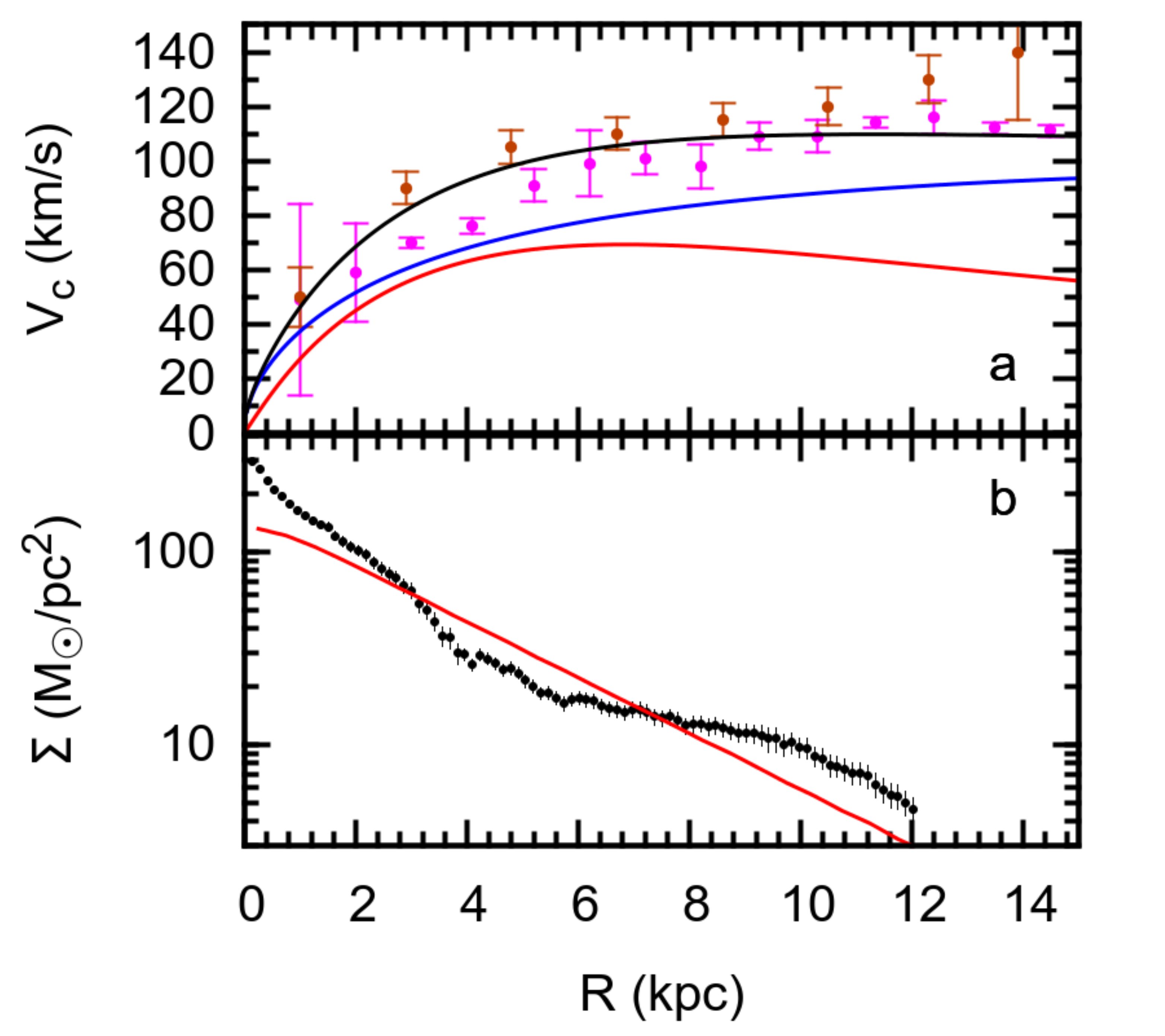}
\caption{The rotation curve and surface density profile of the equilibrium initial conditions for our UGC 628 model. Panel a: Initial disc (red), halo (blue), and total (black) circular velocity curves for the model, and derived rotation curves from \citet{deblokbosma2002} (dark orange points) and \citet{cheminhernandez2009} (magenta points). Panel b: Initial surface density profile (solid red line) overplotted on the inferred UGC 628 profile (black points; see Section~\ref{sec:modellingobservations}). \label{fig:rotcurve_surfdens_ICS}}
\end{figure}

\subsection{Rotation Curve Data}
\label{sec:modellingrotationcurve}

Rotation curve data derived from optical H$\alpha$ velocity field observations from \citet{deblokbosma2002} and \citet{cheminhernandez2009} is shown in Fig.~\ref{fig:rotcurve_surfdens_ICS}a. We see that the two data sets are formally inconsistent with one another, particularly in the range $R \sim$ 3-6 kpc. This discrepancy in the inner region is mitigated when improved H$\alpha$ velocity field data with higher angular sampling, compared to \citet{cheminhernandez2009}, are considered (L.~Chemin,~private~communication). However, we note that neither \citet{cheminhernandez2009} nor \citet{deblokbosma2002} accounted for bisymmetric flows when deriving rotation curves from the measured line-of-sight velocities. Because the bar in UGC 628 projects near the major axis, rotation curves derived ignoring the bar flows are biased low relative to the true circular velocity  \citep{spekkenssellwood2007,dicaireetal2008,randriamampandryetal2016}. We account for this observational bias when comparing our models to the rotation curve data sets in Section~\ref{sec:resultscomparisontougc628}. We do use the general shape of the rotation curve, especially in the outer regions of the galaxy where $V(R)$ is approximately constant, to guide the selection of halo model parameters.

\begin{table} 
\caption{Initial galaxy model parameters. The first block corresponds to parameters used to build the initial equilibrium N-body models using the GalactICS code \citep{kuijkendubinski1995, widrowetal2008}. The second block corresponds to parameters derived from values in the first block.}
\label{tab:initialmodelparameters}
\begin{threeparttable} 
\begin{tabular}{lcc}
\hline
Parameter & Our Model & \citet{deblokbosma2002}\\
\hline
$M_{d}$ [$10^{9}$ $M_{\sun}$]\tnote{a} & 9.239  & 5.882 \\
$R_{d}$ [kpc]\tnote{b} & 2.909  & 4.7 \\
$z_{d}$ [kpc]\tnote{c} & 0.485  & 0.783 \\
$R_{out}$ [kpc]\tnote{d} & 25  & 25 \\
$\sigma_{0}$ [km~s$^{-1}$]\tnote{e} & 31.7 & 10.4 \\
$R_\sigma$ [kpc]\tnote{f} & 2.909  & 4.7 \\
$a_{h}$ [kpc]\tnote{g} & 14 & 10.8 \\
$\sigma_h$ [km~s$^{-1}$]\tnote{h} & 210.0  & 262.5 \\
$r_{200}$ [kpc]\tnote{i} & 112 & 130 \\
\hline
\hline
$Q(2.5R_{d})$~\tnote{j} & 1  & 1 \\
$M_{200}$ [$10^{11}$ $M_{\sun}$]\tnote{k} & 1.843 & 2.807 \\
$c$~\tnote{l} & 8 & 12 \\
$\mathcal{F}_{2.2}^{2}$~\tnote{m} & 0.44 & 0.13 \\
\hline
\hline
\end{tabular}
\begin{tablenotes}
\item Disc mass\tnote{a}; radial disc scale length\tnote{b}; vertical disc scale height\tnote{c}; outer disc truncation radius\tnote{d}; central radial velocity dispersion in the disc\tnote{e}; radial scale length of the radial velocity dispersion profile in the disc\tnote{f}; NFW halo scale length\tnote{g}; central velocity dispersion of the halo\tnote{h}; outer radius of the halo\tnote{i}; Toomre $Q$ parameter at 2.5$R_{d}$~\tnote{j}; halo mass\tnote{k}; halo concentration\tnote{l}; disc mass contribution to the total galaxy mass within $2.2R_d$~\tnote{m}.
\end{tablenotes}
\end{threeparttable}
\end{table}

\begin{figure*} \centering
\includegraphics[angle=0,scale=0.57]{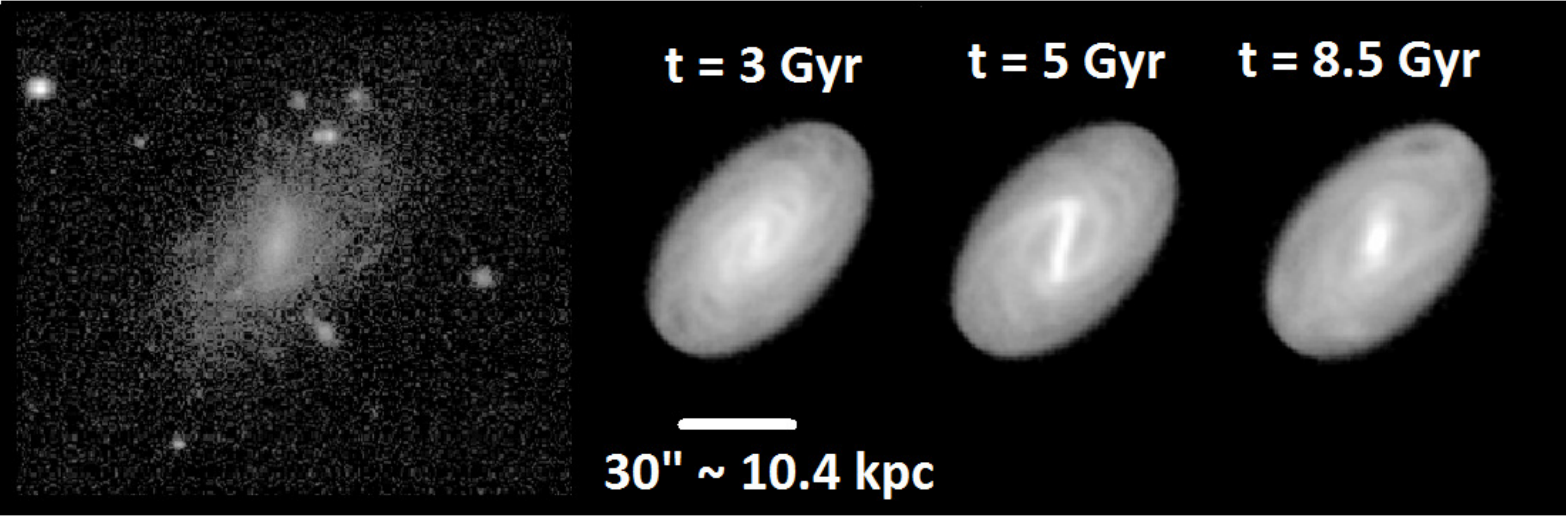}
\caption{The surface brightness distributions of our UGC 628 model when the bar mode begins to grow ($t$ = 3 Gyr), at peak bar strength ($t$ = 5 Gyr), and after buckling ($t$ = 8.5 Gyr), projected on the sky plane using the same log-scale. The leftmost image is a sky-subtracted log-scale $r$-band SDSS image of UGC 628. A distance legend corresponding to both the simulated and observed images is also shown. \label{fig:simskyimage}}
\end{figure*}

\subsection{Initial Conditions for N-body Simulations}
\label{sec:modellingnbody}

We generate initial conditions for our N-body simulations using the
GalactICS code \citep{kuijkendubinski1995, widrowetal2008}, which
allows one to construct axisymmetric, multicomponent galaxy models
that are in approximate dynamical equilibrium.  The models in the
present study comprise a stellar disc and an NFW dark halo.  Particles
for the disc are drawn from a phase space distribution function 
that depends on the energy, angular momentum
about the symmetry axis, and vertical energy.  
By construction, this distribution function yields a disc whose space
density is given by

\begin{equation}
\rho_d(R,\,z) = \frac{M_d}{4\pi R_d^2 \, z_d} e^{-R/R_d} {\rm sech}^2(z/z_d)~,
\end{equation}

\noindent where $M_d$ is the total disc mass and $z_d$ is the disc scale height. In addition, the radial velocity dispersion of the disc is an
exponential function of $R$:

\begin{equation}
\sigma_R(R) = \sigma_{0}e^{-R/2R_\sigma}~,
\end{equation}

\noindent where $\sigma_0$ is the central dispersion and $R_\sigma$ is the radial scale length of the squared dispersion profile. The velocity dispersion in the azimuthal direction is
calculated from the epicycle approximation.  Finally, the disc is
constructed to be approximately isothermal in the direction
perpendicular to the disc plane.
\\
\indent
The halo distribution function depends only on the energy and is
constructed to yield an NFW density profile \citep{navarroetal1996}

\begin{equation}
\rho_h(r) = \frac{a_h\sigma_h^2}{4\pi G}
\frac{1}{r\left (r + a_h\right )^2}~,
\end{equation}

\noindent where $a_h$ is the NFW scale length and $\sigma_h$ is the characteristic velocity scale.
\\
\indent
The input values for $R_d$ and $M_d$ are found by fitting the surface
density profile to a single exponential function, as shown in
Fig.~\ref{fig:rotcurve_surfdens_ICS}b.  We then set $z_d = R_d/6$ as
suggested in a study of edge-on galaxies by
\citet{vanderkruitsearle1981}.  Furthermore, we assume that the
exponential scale length for $\sigma_R^2$ is the same as for the
surface density, i.e. $R_\sigma = R_d$ \citep{bottema1993}.  Finally,
we set the central radial velocity dispersion so that the Toomre $Q$ parameter \citep{toomre1964} is equal to unity
at $R = 2.5R_d$.
\\
\indent
The halo parameters $a_h$ and $\sigma_h$ were chosen to produce a
reasonable fit (chi-by-eye) to the observed outer rotation curve. A comparison of the disc-halo rotation curve decomposition of our initial conditions and the \citet{deblokbosma2002} and \citet{cheminhernandez2009} data is shown in Fig.~\ref{fig:rotcurve_surfdens_ICS}a. If we define the total mass in dark matter by the standard $M_{200}$ (mass interior to radius $r_{200}$, which is defined as the radius inside
which the mean density is 200 times the critical density), then our
model has a disc mass fraction of $f_d = M_d/M_{200}=0.05$ and a halo
concentration $c = 8$.  As noted by \citet{mayerwadsley2004}, LSB
galaxies typically have mass fractions $f_{d}\leq 0.1$ and can be bar
unstable with halo concentrations as low as $c = 4$.
\\
\indent
Parameter values used for our model are presented in
Table~\ref{tab:initialmodelparameters}. The disc and halo
distribution functions were populated with $10^{6}$ and $2 \times
10^{6}$ particles, respectively. The models were simulated using
\textsc{gadget}-2 \citep{springel2005} for $\sim$11.7 Gyr (roughly 195 dynamical times at $R = 2.2R_d$) with a
softening length of 40 pc for both disc and halo
particles. Minimum and maximum time steps were set to 0.01\% and 0.2\%
of the galactic dynamical time defined at a radius of 20 kpc. Energy
was conserved to within a maximum of 0.04\% over the simulation
runtime.
\\
\indent
Additional simulations were run to test the sensitivity of our results
to the model parameters; and we have verified that models with initial parameter values of $2 <
R_{d} < 3.5$ kpc and $1 < Q(2.5R_d) < 1.5$ produce similar
results.  The same cannot be said for an N-body model with the shallower, less dense disc of \citet{deblokbosma2002}.  The
parameters for our realization of their model are also given in
Table~\ref{tab:initialmodelparameters}.  In this case, 
the galaxy develops flocculent spiral structure but
never forms a bar.

\section{Bar Formation and Evolution in the Simulated Models}
\label{sec:results}

\subsection{Comparison to UGC 628}
\label{sec:resultscomparisontougc628}

\begin{figure*} \centering
\includegraphics[angle=0,scale=0.18]{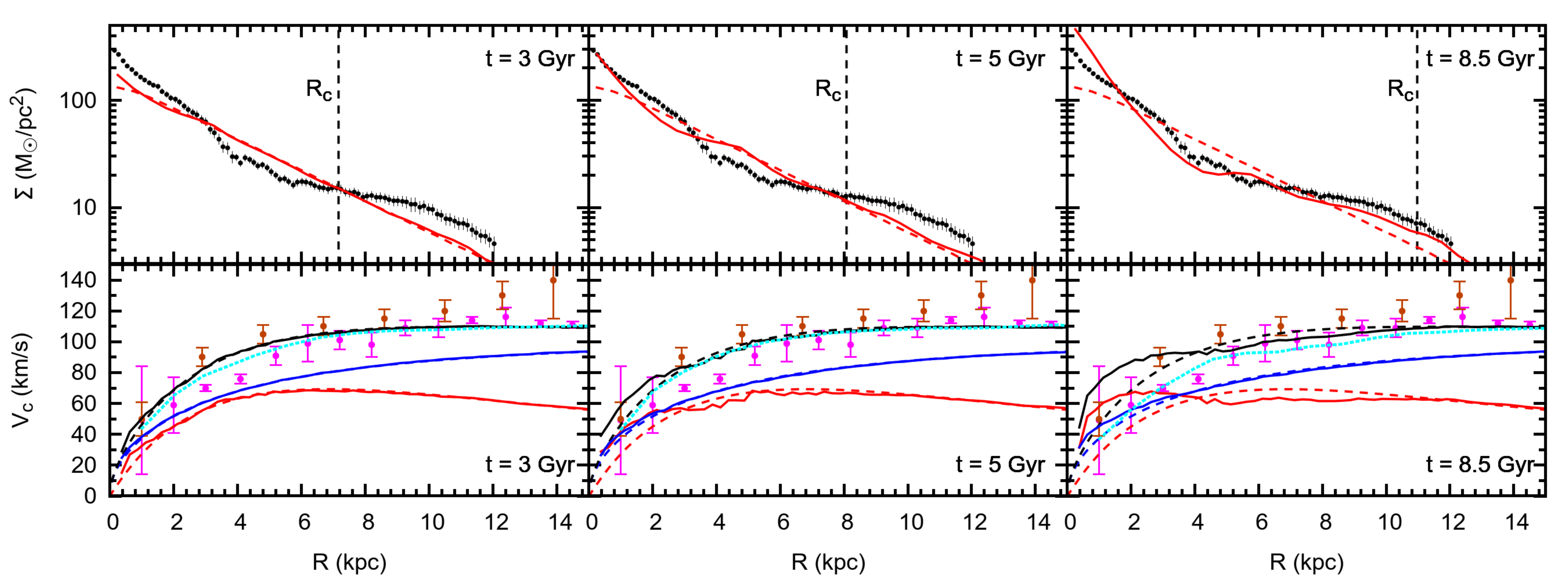}
\caption{Evolution of the mass distribution and rotation curve in our model of UGC 628. Columns correspond to various times in the simulation as indicated. Top row: Surface density profile. The observed surface density profile is shown as black points, the initial conditions of the model are shown as a red dashed line, and the surface density profile of the model at the indicated time is shown as a solid red line. Only uncertainties larger than the data points are shown. Vertical black dashed lines indicate the radius of corotation at each time. Bottom row: Circular rotation curve. The red, blue, and black lines correspond the disc, halo, and total rotation curves, respectively. The model initial conditions correspond to the dashed lines, while the solid lines show the rotation curve at the time indicated. Observational data from \citet{deblokbosma2002} and \citet{cheminhernandez2009} are shown as dark orange and magenta points, respectively. The cyan line shows the rotation curve derived from the line-of-sight velocities for the model projections shown in Fig.~\ref{fig:simskyimage}, under the assumption of axisymmetry, for comparison with the observational data. \label{fig:rotcurve_surfdens_latertimes}}
\end{figure*}

In Fig.~\ref{fig:simskyimage} we show a sky subtracted log-scale $r$-band image of UGC 628 alongside the projected log-scale disc surface brightness from our simulation when the bar mode begins to grow ($t = 3$ Gyr), at peak bar strength ($t = 5$ Gyr), and after buckling ($t = 8.5$ Gyr; see Section~\ref{sec:resultsbarproperties}). The surface brightness images were constructed by applying the mass-to-light fit (Fig.~\ref{fig:observationdata}c) to convert N-body particle mass to luminosity. The bar position angle in the simulations was chosen to match that in the observations. Additionally, the simulated discs were truncated at $R =$ 12.5 kpc before they were projected onto the sky to mimic the sky subtraction used in producing the $r$-band image.
\\
\indent
At all times, the inner disc of our model has a clear central bar. In addition, spiral arms emanating from the ends of the bar are visible, especially in the $t = 3$ and $t = 5$ Gyr snapshots. The bar at 5 Gyr is far narrower than that in the $r$-band image. Conversely, the bar at 8.5 Gyr does not extend quite as far in length as the bar in UGC 628. Thus, the bar in UGC~628 may be best-fit by our model at a time between peak strength and buckling.
\\
\indent
The surface density and rotation curve decomposition profiles at $t =$ 3, 5, and 8.5 Gyr are presented in Fig.~\ref{fig:rotcurve_surfdens_latertimes}. The evolution of the surface density profile in the top row of Fig.~\ref{fig:rotcurve_surfdens_latertimes} shows considerable mass redistribution within the disc. Disc material initially at $R \sim$ 5 kpc is mainly redistributed to the central bar region, although some disc material moves to the outer disc near $R_{c}$. We note that, even just after the formation of the bar at $t = 3$ Gyr, the disc surface density at $R_c$ is low and the shear is high because the rotation curve is already flat.
\\
\indent
The surface density profile at $t = 8.5$ Gyr matches the observed profile decently, and the break in the profile occurs around the correct radius. Additionally, the surface density in the outer disc captures the proper decrease in density apparent in the observational data. Of course, choosing to compare the surface density profile at $t = 8.5$ Gyr to observations is arbitrary, and we find slightly better or worse agreement depending on the simulation snapshot. However, after the bar buckles the surface density stays roughly constant while the disc is continually being heated in both the in-plane and vertical directions.
\\
\indent
The circular velocity curves in the bottom row of Fig.~\ref{fig:rotcurve_surfdens_latertimes} were computed by calculating the total in-plane force from all particles (black line) and from the disc (red line) and halo (blue line) potentials using direct summation on 5000 test particles uniformly distributed within the disc mid-plane. For $R >$ 5 kpc, the model rotation curves are in good agreement with the observed rotation curves at all times. At later times once the bar has buckled, the disc contribution in the bar region is strongly increased. For $R <$ 5 kpc, the initial circular velocity is considerably larger than that measured by \citet{cheminhernandez2009}, but agrees with that obtained by \citet{deblokbosma2002}. The increased circular velocity of the system at later times in the simulation is also broadly consistent with the \citet{deblokbosma2002} measurements except at $R =$ 1 kpc, where the simulated circular velocity exceeds the measured value. As discussed above, this discrepancy between our model and the data is not surprising because \citet{cheminhernandez2009} and \citet{deblokbosma2002} do not account for bisymmetric flows when deriving the rotation curves (see Section~\ref{sec:modellingrotationcurve}). The cyan lines in the bottom panels of Fig.~\ref{fig:rotcurve_surfdens_latertimes} show the rotation curves that we derive from line-of-sight velocities of the model projections in Fig.~\ref{fig:simskyimage}, under the assumption of axisymmetry. There is good agreement between these curves and the observational data. 

\begin{figure} \centering
\includegraphics[angle=0,scale=0.26]{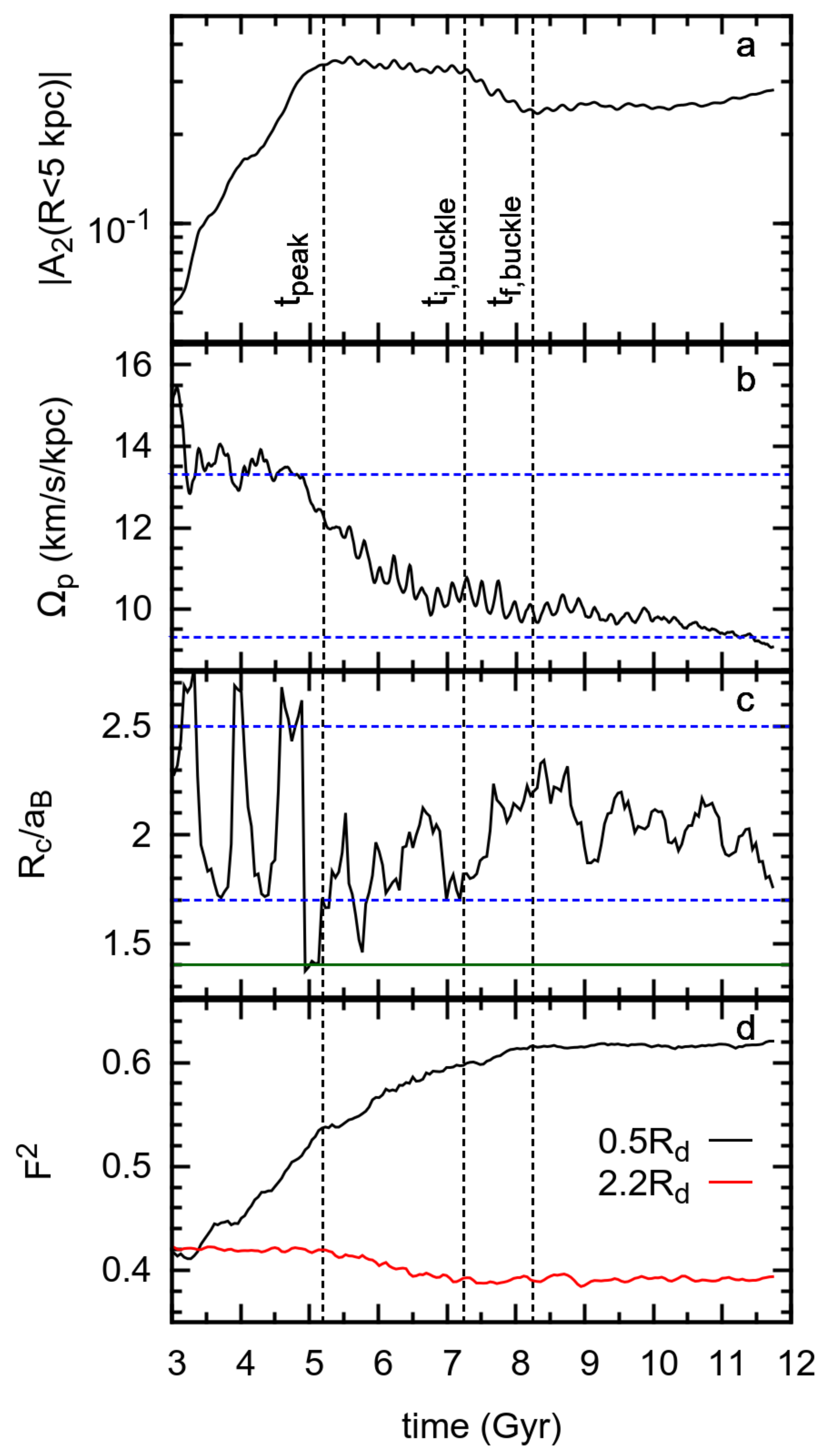}
\caption{Quantitative properties of the bar and disc as a function of time for the UGC 628 model. All quantities are averaged over a galactic dynamical time defined at a radius of 20 kpc. Vertical black dashed lines indicate times of peak strength, start of buckling, and the end of buckling, as labelled. Panel a: Bar strength. Panel b: Bar pattern speed. The range of measured values allowed by the 1$\sigma$ uncertainties of \citet{cheminhernandez2009} is indicated as horizontal blue dashed lines. Panel c: Ratio of corotation radius and bar semi-major axis length. The range of measured values allowed by the 1$\sigma$ uncertainties of \citet{cheminhernandez2009} are indicated as horizontal blue dashed lines. The horizontal solid green line indicates the cut-off between fast and slow bars, $\mathcal{R}$ = 1.4. Panel d: Disk mass fraction. The black and red lines correspond to measurements evaluated at 0.5$R_{d}$ and 2.2$R_{d}$, respectively.  \label{fig:barpropertiesvstime}}
\end{figure}

\begin{figure} \centering
\includegraphics[angle=0,scale=0.25]{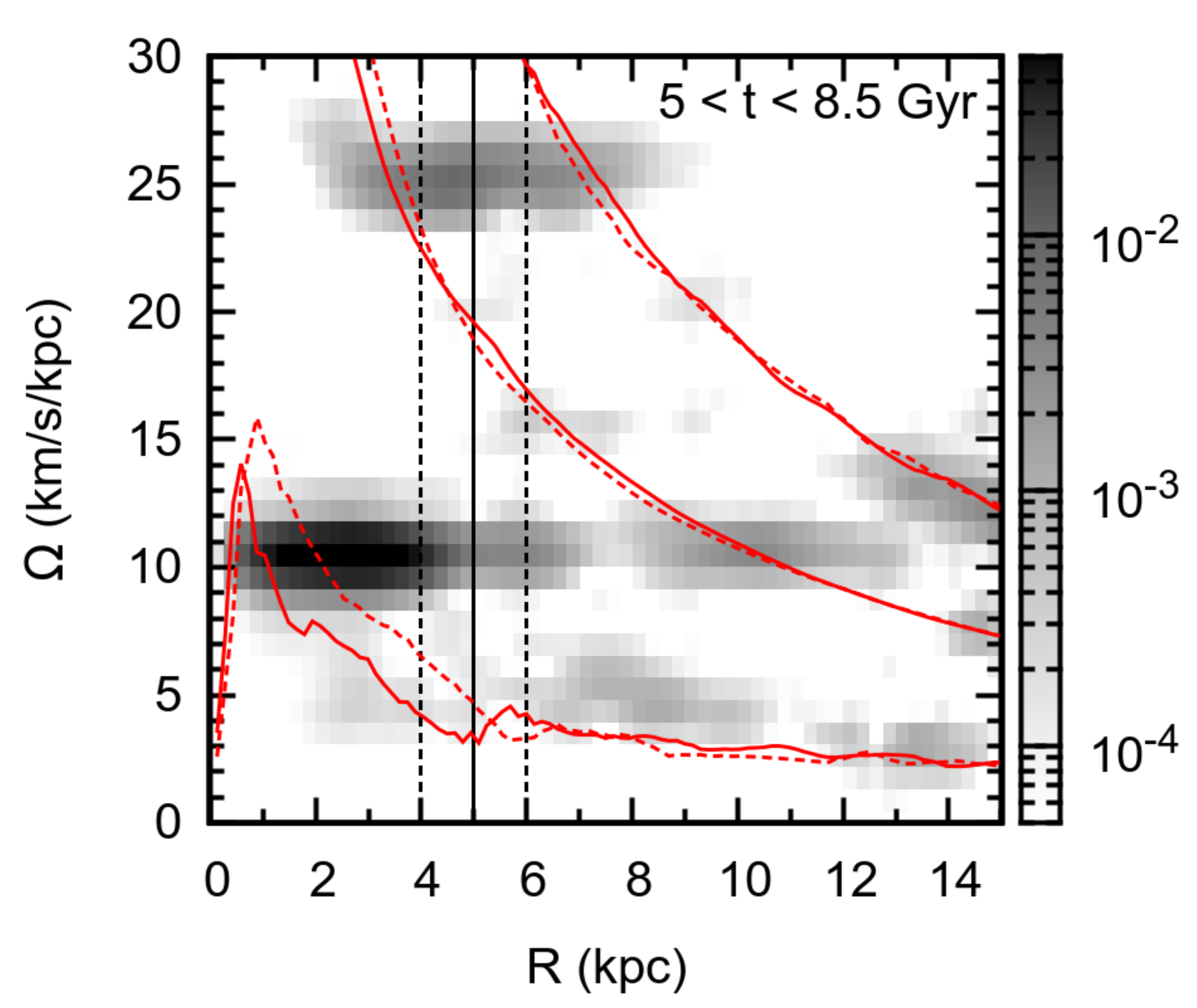}
\caption{Power spectrum of $m$=2 density mode frequencies in the simulated model as a function of cylindrical radius for $5~<~t~<~8.5$~Gyr. The vertical black lines correspond to the average measurement of $a_{B}$ over the specified time frame (solid) and the associated 1$\sigma$ uncertainty (dashed). The red lines show the inner Lindblad, corotation, and outer Lindblad resonances (from left to right) at $t =$ 5 Gyr (solid) and 8.5 Gyr (dashed). \label{fig:fourierspectral}}
\end{figure}

\subsection{Bar Properties}
\label{sec:resultsbarproperties}

Section~\ref{sec:resultscomparisontougc628} established good qualitative agreement between the photometry and kinematics of UGC 628 and those of our barred galaxy model around the time of buckling. To quantify the strength of the bar modes in our model we expand the surface density of the disc into a Fourier series in azimuthal angle $\phi$ of $m$-fold density symmetries and find the magnitude of the $m$=2 Fourier coefficient \citep{sellwoodathanassoula1986} for all disc particles with cylindrical radius $R <$ 5 kpc, the region that encompasses the bar at late times. We assume that the bar is the only bisymmetric mode present for $R <$ 5 kpc at all times, and that contamination from bisymmetric spiral modes is limited. We define the Fourier coefficient for the $m$th mode at time $t$ as

\begin{equation}
A_{m}(t) = \frac{1}{N} \sum\limits_{j=1}^{N} e^{i m \phi_{j}}, \label{eq:fouriercoef}
\end{equation} 

\noindent
where $N$ is the number of particles with $R <$ 5 kpc, $m$=2 for a bisymmetric bar mode, and $\phi_{j}$ is the azimuthal position of the $j$th particle. We define the bar strength as the magnitude of equation~(\ref{eq:fouriercoef}).
\\
\indent
The bar strength as a function of time for our model is shown in Fig.~\ref{fig:barpropertiesvstime}a. After buckling, the bar settles in strength for the remainder of the simulation. In comparison, the bisymmetric mode strength in the constant mass-to-light \citet{deblokbosma2002} model remained at the initial noise level of $|A_{2}| \sim$$10^{-2}$ over the entire simulation runtime, which is more than an order of magnitude smaller than the bar strength measured in any of the variable mass-to-light models that we simulated.
\\
\indent
To compute the bar pattern speed $\Omega_p$, we evaluate the numerical derivative of the cumulative azimuthal phase of the $m$=2 density mode in the inner disc \citep{ridders1982,press2007}. The result is presented in Fig.~\ref{fig:barpropertiesvstime}b and is generally consistent with measurements by \citet{cheminhernandez2009} over the lifetime of the bar, particularly after the time of peak strength at $t \sim$5 Gyr. The pattern speed begins to decrease once the bar reaches maximum strength, presumably due to dynamical friction from the halo \citep{chandrasekhar1943,mulder1983,weinberg1985}. At the onset of bar buckling ($t \sim$7 Gyr) the rate at which the pattern speed decreases changes and stays constant for the remainder of the simulation.
\\
\indent
To quantify the relative speed of the bar we compute $\mathcal{R}$ as a function of time. The radius at which $\Omega(R) = \Omega_{p}$ in Fig.~\ref{fig:barpropertiesvstime}b defines $R_{c}$. Accurate and robust determination of bar lengths in simulations is a long standing challenge \citep{michel-dansacwozniak2006}. We estimate the bar semi-major axis using a method similar to that described in \citet{debattistasellwood2000}, where $a_{B}$ is defined as the location of the upturn in the radial profile of $|A_{2}|$ near the bar edge. As noted by \citet{debattistasellwood2000}, this method tends to underestimate the length of the bar, particularly in the presence of noise. We mitigate this effect by considering the weighted mean of $\mathcal{R}$ over a dynamical time, where outlying values are down-weighted. 
\\
\indent
The resulting values of $\mathcal{R}$ are shown in Fig.~\ref{fig:barpropertiesvstime}c, and fall within the confidence bounds of measurements by \citet{cheminhernandez2009} for most of the simulation once the bar has formed, and especially at later times after the bar has begun to buckle. The large variability in $\mathcal{R}$ relative to $\Omega_{p}$ and $|A_{2}|$ (Fig.~\ref{fig:barpropertiesvstime}a,b) arises because it is a ratio of two noisy quantities, $R_{c}$ and $a_{B}$, particularly for $t <$ 5 Gyr when the bar is still growing. It is clear that the bar in our model is slow, with $\mathcal{R} >$ 1.4 over the entire bar lifetime. We note that both $R_c$ and $a_B$ do increase over time, however, $a_B$ does so at a lesser rate and would have to approximately double in value for the bar to be considered ``fast".
\\
\indent
A more visual depiction of the bar's relative speed is shown in Fig.~\ref{fig:fourierspectral}, where the power spectrum of bisymmetric mode frequencies is plotted as a function of radius for $5 < t < 8.5$ Gyr \citep[see also \citet{roskaretal2012} for a succinct description]{sellwoodathanassoula1986,press2007}. The epicyclic frequency used to compute resonant frequencies in Fig.~\ref{fig:fourierspectral} was calculated from

\begin{equation}
\kappa^2 = 4 \Omega^{2}(R) + R \frac{d \;}{dR} \Omega^{2}(R),
\end{equation}

\noindent
where $\Omega(R) = V(R)/R$ is the circular frequency profile of the galaxy and $V(R)$ is the circular speed computed from the in-plane forces, as described in Section~\ref{sec:resultscomparisontougc628} \citep{binneytremaine2008}. The peak in power at $(R, \Omega) \sim$ (3 kpc, 10 $\textrm{km} \, \textrm{s}^{-1} \, \textrm{kpc}^{-1}$) corresponds to the bar. The structure at $\Omega \sim$ 25 $\textrm{km} \, \textrm{s}^{-1} \, \textrm{kpc}^{-1}$ between corotation and the outer Lindblad resonance corresponds to a bisymmetric spiral mode emanating from the edge of the bar. Also present in Fig.~\ref{fig:fourierspectral} are bisymmetric density modes that have accumulated along various resonances. The bar structure clearly does not extend to corotation, which is apparent from the order of magnitude difference in power between the peak power at $R \sim$ 3 kpc and the structure closer to corotation along $\Omega \sim$ 10 $\textrm{km} \, \textrm{s}^{-1} \, \textrm{kpc}^{-1}$. Thus, the bar is slow since it is rotating at a frequency far less than $\Omega(a_{B})$.
\\
\indent
To investigate how the relative disc and halo mass distributions change as a function of time we compute $\mathcal{F}^{2}$, from equation~(\ref{eq:discmaximality}), as a function of time evaluated at various multiples of the disc radial scale length of an exponential fit to the radial surface density profile for $R <$ 12 kpc \citep{press2007}. Fig.~\ref{fig:barpropertiesvstime}d shows $\mathcal{F}^{2}_{0.5}$ and $\mathcal{F}^{2}_{2.2}$ as a function of time, which serves as proxies for the mass contribution of the disc within $R = 0.5R_d$ and $R = 2.2R_d$, respectively. When the bar starts to form, $\mathcal{F}^{2}_{0.5} \sim \mathcal{F}^{2}_{2.2} \sim$ 0.4. By the time the bar reaches peak strength there are equal mass fractions between the disc and halo in the bar region. The disc mass contribution steadily increases over time in the inner region and slightly decreases within $ R = 2.2R_{d}$ until the bar buckles. After buckling, the relative fraction of disc and halo mass remains roughly constant in time with $\mathcal{F}^{2}_{0.5} \sim$ 0.6 and $\mathcal{F}^{2}_{2.2} \sim$ 0.4. Thus, between the times of peak bar strength and on the onset of buckling, when the model best corresponds to the photometry and kinematics of UGC~628, the model is baryon dominated in the bar region and dark matter dominated farther out.

\section{Discussion and Summary}
\label{sec:discussion}

We have presented N-body models of the barred LSB galaxy UGC~628, one of the few systems for which a slow ($\mathcal{R} = R_c/a_B > 1.4$) bar has been measured \citep{cheminhernandez2009}. We re-examined the surface brightness profile of UGC~628 by fitting SDSS DR10 photometry and applying the colour-mass-to-light ratio transformations of \citet{intoportinari2013}, finding a higher surface density than previously reported \citep{deblokbosma2002}. We used this surface density distribution together with rotation curve data from the literature to initialize our simulations with an exponential stellar disc with a fractional mass contribution enclosed within $R=2.2R_d$ of $\mathcal{F}_{2.2}^{2} = 0.44$  embedded in an NFW halo. The disc developed a bar that begins to grow at $t=3\,$Gyr, peaks in strength at $t=5\,$Gyr, and finishes buckling by $t=8.5\,$Gyr. 
\\
\indent
The model provides a good description of the available photometry and kinematics of UGC~628 at relatively late times, between the bar's peak strength and buckling. The bar length and pattern speed that we measure from the simulations imply $\mathcal{R} \sim 2 $ after the bar reaches peak strength, in broad agreement with the measurements of \citet{cheminhernandez2009}. We find that the bar redistributes mass effectively in the inner disc; our model therefore implies that UGC~628 is baryon dominated in the bar region and dark matter dominated further out.
\\
\indent
The observations for UGC~628 by \citet{cheminhernandez2009} and the models presented in this paper imply that, in agreement with the photometric models of \citet{rautiainenetal2008}, at least some late-type galaxies host slow bars. The pattern speed evolution of the bar in our model is broadly consistent with theoretical considerations and N-body simulations of submaximal discs, where drag from dynamical friction decreases the pattern speed \citep[see][for a review]{sellwood2014}. However, the bars in most simulations are ``born fast" with  $\mathcal{R} \sim 1$ just after they form \citep[e.g.][]{debattistasellwood2000}, while $\mathcal{R} \sim 2$ in our model at all times. We attribute this difference to the relatively low stellar surface density and high shear in our model at $R=R_c$, which follow from the LSB classification and flat rotation curve of UGC~628, even at early times (Fig.~\ref{fig:rotcurve_surfdens_latertimes}): the disc therefore does not have sufficient self-gravity to support a bar mode out to corotation even before any braking has taken place.
\\
\indent
Our model has a disc mass fraction within 2.2 exponential scale lengths of $\mathcal{F}^{2}_{2.2} \sim 0.4$ throughout the simulation (Fig.~\ref{fig:barpropertiesvstime}d), and is therefore dark matter dominated according to that definition. With $V_{tot} = 110 \, \mathrm{km\,s^{-1}}$ and $f_{DM} = 1 - \mathcal{F}^{2}_{2.2} \sim 0.6$, our model sits on the lower $1\sigma$ envelope of the universal relation proposed by \citet{courteaudutton2015}, commensurate with LSB galaxies having systematically larger discs than their high surface brightness counterparts \citep[e.g.][]{zwaanetal1995}. Contrary to the conclusion of \citet{cheminhernandez2009} given the low pattern speed of the bar, however, our model suggests that UGC~628 is not dark matter dominated at all radii but instead has $\mathcal{F}^{2}_{0.5} \sim 0.6$ at the onset of bar buckling. Thus our model of UGC~628 provides an example of the fact that, as argued by many authors in the past \citep{debattistasellwood2000,courteaudutton2015}, the disc mass fraction enclosed within the radius where its contribution peaks is not a reliable indicator of its dynamical importance at all radii. Pairing $\mathcal{F}^{2}_{2.2}$ with a measure at smaller radii, such as $\mathcal{F}^{2}_{0.5}$, may be more informative (Fig.~\ref{fig:barpropertiesvstime}d), although the latter is even harder to constrain observationally than the former due to uncertainties in stellar population models, dust obscuration corrections, and bulge contributions near galaxy centres, particularly in more massive systems than the one considered here. 
\\
\indent 
We note that simulations show that the presence of gas during bar growth and evolution can influence its final properties \citep[e.g.][but see \citealt{sellwooddebattista2014}]{bournaudetal2005,villavargasetal2010,athanassoula+2013,athanassoula2014}. Most of these simulations distribute the gas identically to the stars at the outset, and therefore reflect the influence of molecular gas discs rather than the more extended atomic gas ones \citep[e.g.][]{broeilsrhee1997,franketal2016}. The low star formation efficiency and quiescent star formation history of UGC~628 \citep{youngetal2015} suggest a low molecular-to-atomic gas ratio both now and in the past, consistent with CO non-detections in late-type LSB galaxies \citep{das2006}.  In addition, UGC~628 itself has an unusually low atomic-to-stellar mass ratio of $M_{HI}/M_* \sim 0.2$ for an LSB galaxy \citep{springobetal2005, kim2007}. We therefore expect the molecular gas fraction in UGC~628 to be low and its inclusion in our models to have little influence on the bar properties that we report here \citep{berentzenetal2007}. 
\\
\indent
The baryon dominated central regions of UGC~628 implied by our model distinguish it from most others LSB galaxies, where kinematics and photometry suggest a dark matter dominated inner disc \citep{bothunetal1997,deblokmcgaugh1997,debloketal2001,combes2002,deblokbosma2002,kuziodenarayetal2008}. The disc in UGC~628 therefore has enough self-gravity to support a bar but also a sufficiently low central surface brightness ($\mu_{B,0} = 23.1\,\mathrm{mag\,arcsec^{-2}}$, \citealt{kim2007}) for the system to be classified as an LSB. Indeed, the rarity of bars in LSB galaxies \citep{mihosetal1997} suggests that this balance is a delicate one. It may be the case that the properties of the bar and mass distribution in UGC~628 found here also apply to the few other known barred LSB galaxies, though to our knowledge pattern speeds for these systems have not been measured. Estimates of $\mathcal{R}$ and detailed dynamical models for these LSB galaxies as well as the other known systems with slow bars \citep{bureauetal1999,banerjeeetal2013} may help further explore the properties of such instabilities in late-type, low-mass, and LSB discs.

\section*{Acknowledgements}
We would like to thank the anonymous referee for providing suggestions that improved the quality of this paper,  E. Athanassoula for useful conversations, L. Chemin for generously providing data for a graduate class project from which this study emerged, and N. Deg for the algorithm used to produce sky projections of the simulated model in Fig.~\ref{fig:simskyimage}. We also acknowledge the use of computational resources at the Centre for Advanced Computing.  MHC acknowledges the financial support of the Ontario Graduate Scholarship, Queen Elizabeth II Graduate Scholarship in Science and Technology, and Natural Sciences and Engineering Research Council Postgraduate Scholarship programmes throughout the duration of this work. KS and LMW are supported by the Natural Sciences and Engineering Research Council of Canada through Discovery Grants. We acknowledge the use of data from SDSS-III. Funding for SDSS-III has been provided by the Alfred P. Sloan Foundation, the Participating Institutions, the National Science Foundation, and the U.S. Department of Energy Office of Science.




\bibliographystyle{mnras}
\bibliography{bibliography}

\bsp	
\label{lastpage}
\end{document}